\documentclass[11pt,a4paper]{article}
\usepackage{jcappub}
\usepackage{euscript,epsfig,amsmath,amssymb}
\usepackage{amsfonts,latexsym}
\usepackage{array}
\usepackage{enumerate}
\usepackage{booktabs}
\usepackage{tabularx}

\usepackage{hyperref}
\usepackage{url}
\usepackage[normalem]{ulem}

\newcolumntype{L}[1]{>{\hsize=#1\hsize\raggedright\arraybackslash}X}%
\newcolumntype{R}[1]{>{\hsize=#1\hsize\raggedleft\arraybackslash}X}%
\newcolumntype{C}[1]{>{\hsize=#1\hsize\centering\arraybackslash}X}%

\newcommand{\be}[1]{\begin{equation}\label{#1}}
\newcommand{\ee}{\end{equation}}
\newcommand{\ba}[1]{\begin{eqnarray}\label{#1}}
\newcommand{\ea}{\end{eqnarray}}
\newcommand{\rf}[1]{(\ref{#1})}
\newcommand{\nn}{\nonumber}

\newcommand{\const}{\mbox{\rm const}\,}

\newcommand{\cH}{\mathcal{H}}							
\newcommand{\cK}{\mathcal{K}}							

\newcommand{\epsdm}{\overline \varepsilon_{\mathrm{DM}}}	
\newcommand{\epsde}{\overline \varepsilon_{\mathrm{DE}}}	

\newcommand{\epsdmnow}{\overline \varepsilon_{\mathrm{DM},0}}	
\newcommand{\epsdenow}{\overline \varepsilon_{\mathrm{DE},0}}	

\newcommand{\delepsrad}{\delta\varepsilon_{\mathrm{rad}}}	
\newcommand{\delepsdm}{\delta \varepsilon_{\mathrm{DM}}}	
\newcommand{\delepsde}{\delta \varepsilon_{\mathrm{DE}}}	
\newcommand{\dQ}{{\delta Q}}							

\begin{document}

\title{The late Universe with non-linear interaction in the dark sector: the coincidence problem}

\author{Mariam Bouhmadi--L\'opez$^{1,2,3,4}$,}
\author{Jo\~ao Morais$^{3}$,}
\author{and Alexander~Zhuk$^{5}$,}

\affiliation{
${}^1$ Departamento de F\'isica, Universidade da Beira Interior, 6200 Covilh\~a, Portugal \\
${}^2$ Centro de Matem\'atica e Aplica\c{c}\~oes da Universidade da Beira Interior (CMA-UBI), 6200 Covilh\~a, Portugal\\
${}^3$ Department of Theoretical Physics, University of the Basque Country
UPV/EHU, P.O. Box 644, 48080 Bilbao, Spain\\
${}^4$IKERBASQUE, Basque Foundation for Science, 48011, Bilbao, Spain\\
$^{5}$Astronomical Observatory, Odessa National University,\\ Dvoryanskaya st. 2, Odessa 65082, Ukraine\\
}

\emailAdd{mbl@ubi.pt (On leave of absence from UPV/EHU and IKERBASQUE)}
\emailAdd{jviegas001@ikasle.ehu.eus}
\emailAdd{ai.zhuk2@gmail.com}

\abstract{
We study the Universe at the late stage of its evolution and deep inside the cell of uniformity. At such a scale the Universe is highly inhomogeneous and filled with discretely distributed inhomogeneities in the form of galaxies and groups of galaxies. As a matter source, we consider dark matter (DM) and dark energy (DE) with a non-linear interaction $Q = 3\mathcal{H}\gamma \overline\varepsilon_{\mathrm{DE}} \overline\varepsilon_{\mathrm{DM}}/(\overline\varepsilon_{\mathrm{DE}} + \overline\varepsilon_{\mathrm{DM}})$, where $\gamma$ is a constant. We assume that DM is pressureless and DE has a constant equation of state parameter $w$. In the considered model, the energy densities of the dark sector components present a scaling behaviour with $\overline\varepsilon_{\mathrm{DM}}/\overline\varepsilon_{\mathrm{DE}} \sim \left({a_0}/{a}\right)^{-3(w+\gamma)}$. We investigate the possibility that the perturbations of DM and DE, which are interacting among themselves, could be coupled to the galaxies with the former being concentrated around them. To carry our analysis, we consider the theory of scalar perturbations (within the mechanical approach), and obtain the sets of parameters $(w,\gamma)$ which do not contradict it. We conclude that two sets: $(w=-2/3,\gamma=1/3)$ and $(w=-1,\gamma=1/3)$ are of special interest. First, the energy densities of DM and DE on these cases are concentrated around galaxies confirming that they are coupled fluids. Second, we show that for both of them, the coincidence problem is less severe than in the standard $\Lambda$CDM. Third, the set $(w=-1,\gamma=1/3)$ is within the observational constraints. Finally, we also obtain an expression for the gravitational potential in the considered model.
}

\maketitle

\flushbottom


\section{Introduction}

\

Recent observations \cite{Planck} clearly indicate that our Universe is dark. The mass-energy balance of the Universe consists of 69\% of dark energy (DE) and 26\% of dark matter (DM). However, the nature of these constituents is still unclear. In addition, the $\Lambda$CDM model (as well as a lot of other dark energy models) faces the coincidence problem, i.e. why is the cosmological constant at present of the same order of magnitude as the dark matter energy density? To solve these problems, different dynamical dark energy models were proposed. Among them, the models with interacting DE and DM are of great interest. A vast literature is devoted to these models, their implication in the future evolution of the universe, and their compatibility with the current observational data (see, e.g., \cite{23a,ABZ,Bolotin,Salvatelli:2014zta,Valiviita:2015dfa,Abdalla:2014cla} and numerous references there). These interacting models can be roughly split into two classes: the models with linear and non-linear interaction, respectively. For the former models, the interaction term is proportional to either the energy density of DM, $\overline\varepsilon_{\mathrm{DM}}$, the energy density of DE, $\overline\varepsilon_{\mathrm{DE}}$, or their linear combinations. However, as it was noted in \cite{observ rectr}, an interaction term proportional to $\overline\varepsilon_{\mathrm{DM}}$ leads to a large-scale instability at early times%
\footnote{It is worth noting that such large-scale instability problem can be  solved within a generalized parametrized post-Friedmann approach \cite{PPF1,PPF2}.}%
, whereas a term proportional to $\overline\varepsilon_{\mathrm{DE}}$ could give rise to a negative $\overline\varepsilon_{\mathrm{DM}}$ in the future. Therefore, models where an interaction term is described by a non-linear function of $\overline\varepsilon_{\mathrm{DM}}$ and $\overline\varepsilon_{\mathrm{DE}}$ were proposed (see, e.g., \cite{23a,ABZ,observ rectr,Magnano,24a,Baldi,25a,Poitras,2Zhang,Oliveros,add1,add2,Xin}). In these papers, in most of the cases the interaction term is a product of the energy densities of DM and DE. This is motivated by the analogy with two-body chemical reactions where the interaction rate is proportional to the product of the number densities of the two particles species \cite{23a,24a}. Following this reasoning, in our paper we consider an interaction term of the form:
$$
	Q = 3\mathcal{H}\gamma \frac{\overline\varepsilon_{\mathrm{DE}} \overline\varepsilon_{\mathrm{DM}}}{\overline\varepsilon_{\mathrm{DE}} + \overline\varepsilon_{\mathrm{DM}}}\,,
$$
where $\gamma$ is a dimensionless parameter that modules the strength of the interaction.
It can be easily seen that such a model is free of the above mentioned shortcomings present in models with a linear interaction in the dark sector \cite{observ rectr}. We also suppose that DM is pressureless and DE satisfies the equation of state (EoS) $\overline p_{\mathrm{DE}}=w\overline\varepsilon_{\mathrm{DE}}$, where $w=\const$. It was shown in \cite{ABZ} that this model satisfies the scaling solution
$$
\xi \equiv \frac{\epsdm}{\epsde}
	= \xi_0 \left(\frac{a_0}{a}\right)^{-3(w+\gamma)}\,,\quad \xi_0=\const\,.
$$
This solution follows from the conservation equations for DM and DE with the above described interaction term. Notice that some papers start directly from the scaling solution ansatz to investigate the coincidence problem%
\footnote{The coincidence problem within interacting models was also discussed in \cite{add3,add4}.}%
 \cite{38a,40a,41a}. Obviously, if $w+\gamma =0$, the ratio $\xi=\xi_0$
is a constant and, consequently, the coincidence problem is absent (if $\xi_0\sim O(1)$). To get the late-time DE dominance, we should demand that $w+\gamma <0$. In the case of the standard $\Lambda$CDM model $w=-1, \gamma=0 \Rightarrow w+\gamma =-1$. Therefore, for any DM-DE interacting model with $|w+\gamma|<1$, the coincidence problem is less severe than in $\Lambda$CDM. In our paper, we will provide two of such models with $|w+\gamma|=1/3,2/3$.

We would like to stress that the quadratic interactions we have assumed is phenomenological:  (i) it was introduced in Ref.~\cite{observ rectr} as a mean to remove the instabilities present at the perturbative level of some linear DE-DM interaction and (ii) is inspired on the two-body chemical reactions where the interaction rate is proportional to the densities of the two type of particles involved \cite{23a,24a}. We would like also to stress that the non-linear interaction we assume, arise in a natural way within the new generalised Chaplygin gas (NGCG) \cite{observ rectr}. This feature might be very appealing as it might hint towards a particle physics motivations for this interaction as the standard Chaplygin gas can be obtained from d-branes physics and has a super-symmetric realisation \cite{Jackiw:2000mm}. Yet, we must say we are not aware that this is necessarily the case for a general NGCG.

In our analysis we consider the Universe at late-time and deep inside the cell of uniformity. At such scales, the Universe is highly inhomogeneous, and inhomogeneities in the form of galaxies and groups of galaxies have non-relativistic peculiar velocities. In this case, the mechanical approach is an adequate tool to study the scalar perturbations \cite{EZcosm1,EZcosm2}. Further methods related to the mechanical approach were also proposed in \cite{Villa,Uemh,Chis,Adamek,Ellis}.
This approach enables us to get the gravitational potential and to consider the motions of galaxies \cite{EKZ2}. The mechanical approach was applied to a number of DE models to study their compatibility with the theory of scalar perturbations. For example, we considered a perfect fluid with a constant equation of state parameter \cite{BUZ1}, the model with quark-gluon nuggets \cite{Laslo2}, the CPL model \cite{CPL}, the Chaplygin gas model \cite{Chaplygin}, the non-linear $f(R)$ model \cite{Novak} and the model with a scalar field \cite{scal field}. One of the main features of all these models is that perfect fluids (as well as scalar field) are considered in a very specific "coupled" form \cite{coupled}. This means that their fluctuations are concentrated around the galaxies, therefore screening their gravitational potential. Consequently, the peculiar velocities of such coupled perfect fluids are also non-relativistic. In the present paper, we investigate the possibility that non-linearly interacting DM and DE are coupled to the galaxies. The linear interacting model within the mechanical approach was considered in \cite{EinClaus}.
In what follows, we shall show that non-linearly interacting DM and DE can be coupled to the galaxies. They possess a number of interesting properties. First, the energy densities of interacting DM and DE are concentrated around the galaxies. Second, the sum of parameters $|w+\gamma|<1$ is less than 1. Hence, the coincidence problem is not so severe as for the $\Lambda$CDM model (see the definition of the ratio $\xi$ above). Third, the set of parameters $(w=-1,\gamma=1/3)$ is within the current cosmological observations \cite{observ rectr}.

Our paper is structured as follows. In section~\ref{Sec2}, we describe the background model for the non-linear interacting DM and DE. In section~\ref{Sec3}, we consider the theory of scalar perturbations for the considered model. Here, we define the sets of parameters which do not contradict the theory of scalar perturbations. In the concluding section~\ref{Sec4}, we summarise the obtained results and discuss them.


\section{Non-linear interaction in the dark sector: background equations}

\label{Sec2}

\setcounter{equation}{0}

\

We start with the Friedmann and Raychaudhuri equations for the homogeneous background which read
\be{2.1} \frac{3\left(\mathcal{H}^2+\mathcal{K}\right)}{a^2}= \kappa\left(\overline{T}^0_{0}+\overline \varepsilon_{\mathrm{rad}} + \overline
\varepsilon_{\mathrm{DM}} + \overline\varepsilon_{\mathrm{DE}} \right) +\Lambda
\ee
and
\be{2.2}
\frac{2\mathcal{H}'+\mathcal{H}^2+\mathcal{K}}{a^2}=-\kappa\left(\overline p_{\mathrm{rad}} + \overline p_{\mathrm{DE}} \right) + \Lambda\,,
\ee
where ${\mathcal H}\equiv a'/a\equiv (da/d\eta)/a$, \ $\kappa\equiv 8\pi G_N/c^4$ ($c$ is the speed of light and $G_N$ is the Newtonian gravitational
constant) and $\mathcal K=-1,0,+1$ for open, flat and closed Universes, respectively.
The conformal time, $\eta$, and the synchronous or cosmic time, $t$, are connected as $cdt=a d\eta$, where $a$ is the scale factor. As usual, for radiation we have the EoS $\overline p_{\mathrm{rad}}=(1/3)\overline \varepsilon_{\mathrm{rad}}$. $\overline T^{0}_{0} =\overline \rho_{\mathrm{c}} c^2/a^3$ is the average energy density of the non-interacting non-relativistic matter with the comoving rest mass density $\overline \rho_{\mathrm{c}}$. In general, we assume that dark matter may exist both as an interacting and a non-interacting fluid. This type of models was recently investigated, e.g. in \cite{Trieste1,Trieste2}. In our case, $\overline\rho_{\mathrm{c}}$ accounts for non-interacting cold dark matter and baryonic matter. The pressure for all kinds of non-relativistic matter (interacting as well as non-interacting) is assumed to vanish, i.e., $\overline p_{\mathrm{DM}}=0$. On the other hand, the equation of state for the interacting dark energy is
\be{2.3}
\overline p_{\mathrm{DE}} = w\; \overline \varepsilon_{\mathrm{DE}}\,, \quad w\neq 0\,.
\ee
Such a perfect fluid may result under some conditions in an accelerating expansion of the Universe. This is the reason to call it "dark energy".
While it is natural to expect that the current accelerated expansion of the universe requires a negative $w$ (in fact, in the absence of interaction the violation of the strong energy condition imposes that $w<-1/3$), in general, we shall not restrict ourself to such values of $w$.
On the other hand, the presence of such a perfect fluid does not guarantee that the acceleration will happen for sure at present.
Therefore, we still keep in Eqs.~\rf{2.1} and \rf{2.2} the cosmological constant term $\Lambda$. It can be easily realized that even though the presence of such a term influences the dynamical behaviour of the Universe at the background level, it will not affect the analysis of the perturbations.
In the case when the interacting dark energy provides the late-time accelerated expansion of the Universe, we can drop $\Lambda$ and investigate this model from the point of its concordance with the observable data. Such models with dynamical dark energy may help to resolve the coincidence problem.

From Eqs.~\rf{2.1} and \rf{2.2} we get the following auxiliary equation:
\be{2.4}
\frac{2}{a^2}\left(\mathcal{H}'-\mathcal{H}^2-\mathcal{K}\right)= -\kappa\left(\overline{T}^0_{0}+\overline
\varepsilon_{\mathrm{rad}} +\overline \varepsilon_{\mathrm{DM}} +\overline \varepsilon_{\mathrm{DE}} + \overline p_{\mathrm{rad}}+ \overline p_{\mathrm{DE}}\right)\,.
\ee
The non-interacting components of matter/energy are conserved independently. This results, e.g., in a $1/a^3$ and $1/a^4$ dependence for the energy densities of non-relativistic matter and radiation, respectively. However, the interacting components should satisfy the common conservation equation:
\be{2.8}
\overline\varepsilon_{\mathrm{tot}}' + 3 \mathcal{H}\left(\overline\varepsilon_{\mathrm{tot}} + \overline p_{\mathrm{tot}}\right)=0\,,
\ee
where in our case $\overline\varepsilon_{\mathrm{tot}}=\overline\varepsilon_{\mathrm{DM}}+\overline\varepsilon_{\mathrm{DE}}$
and $\overline p_{\mathrm{tot}}=\overline p_{\mathrm{DM}}+\overline p_{\mathrm{DE}}=\overline p_{\mathrm{DE}}$. We can split this equation into two:
\be{2.9}
\overline\varepsilon_{\mathrm{DM}}' + 3 \mathcal{H}\left(\overline\varepsilon_{\mathrm{DM}} + \overline p_{\mathrm{DM}}\right)=
\overline\varepsilon_{\mathrm{DM}}' + 3 \mathcal{H}\overline\varepsilon_{\mathrm{DM}} = Q
\,,
\ee
and
\be{2.10}
\overline\varepsilon_{\mathrm{DE}}' + 3 \mathcal{H}\left(\overline\varepsilon_{\mathrm{DE}} + \overline p_{\mathrm{DE}}\right)=
\overline\varepsilon_{\mathrm{DE}}' + 3\left(1+w\right) \mathcal{H}\overline\varepsilon_{\mathrm{DE}} = -Q, \quad w\neq 0
\,.
\ee
Here, the term $Q$ describes the interaction between dark energy (DE) and dark matter (DM). If $Q>0$, there is a transfer of energy from DE to the interacting DM component, and vice versa for $Q<0$. As we have mentioned in the introduction, there are a quite large number of phenomenological expressions for $Q$. In the present paper, we investigate the case of the non-linear interaction \cite{ABZ,observ rectr}:
\be{2.11}
Q = 3\mathcal{H}\gamma \frac{\overline\varepsilon_{\mathrm{DE}} \overline\varepsilon_{\mathrm{DM}}}{\overline\varepsilon_{\mathrm{tot}}}\,,
\qquad \overline\varepsilon_{\mathrm{tot}}=\overline\varepsilon_{\mathrm{DM}}+\overline\varepsilon_{\mathrm{DE}}\,,
\ee
where $\gamma = \const$ is a free parameter of the model. Since we consider an expanding Universe, we have $\mathcal{H}>0$ at all times, and consequently $\mathrm{sign}\, Q = \mathrm{sign}\, \gamma$.
If the interaction is switched off, i.e., if we set $\gamma =0$, it follows immediately from Eqs.~\rf{2.9} and \rf{2.10} that:
\be{2.11a}
\overline\varepsilon_{\mathrm{DM}} \propto \left(\frac{a_0}{a}\right)^3\,, \qquad
\overline\varepsilon_{\mathrm{DE}} \propto \left(\frac{a_0}{a}\right)^{3(1+w)}\,.
\ee

As found in \cite{ABZ}, the system of Eqs.~\rf{2.9} and \rf{2.10} with the interaction \rf{2.11} is analytically solvable. First, it was shown that the ratio $\xi = \epsdm/\epsde$ for the considered model satisfies the scaling solution:
\be{0.4}
	\xi = \frac{\epsdm}{\epsde}
	= \xi_0 \left(\frac{a_0}{a}\right)^{-3(w+\gamma)}
	\,,
\ee
where $\xi_0\equiv\epsdmnow/\epsdenow$. It is worth noting that cosmological models with the scaling ansatz \rf{0.4} were also investigated in \cite{38a,40a,41a}.
In addition, we can re-write the conservation Eqs.~\eqref{2.9} and \eqref{2.10} as
\ba{0.5}
	&&\epsdm' + 3 \cH\left(1 - \frac{\gamma}{1+\xi}\right)\epsdm =0
	\,,
	\\
	\label{0.6}
	&&\epsde' + 3\cH\left(1 + w + \gamma\frac{\xi}{1+\xi}\right) \epsde =0
	\,.
\ea
From these equations we can identify the effective EoS parameters for the interacting DM and DE
\ba{0.5a}
	w_\textrm{DM}^\textrm{(eff)} = -\frac{\gamma}{1+\xi}
	\,,
	\qquad
	w_\textrm{DE}^\textrm{(eff)} = w+\frac{\gamma\xi}{1+\xi}
	\,.
\ea
There are three possible scenarios for the background evolution, cf. equation~\eqref{0.4}, depending on the sign of $w+\gamma$. We now analyse each case individually.

\

\subsection{$w+\gamma>0$}

	\
	
In this case DM becomes dominant at late-times, since $\xi(a\rightarrow0)=0$ and $\xi(a\rightarrow+\infty)=+\infty$. Therefore, it is of no interest to cosmology with late-time DE dominance.
\\
\subsection{$w+\gamma=0$}
\

This leads to scaling solutions (self-similar solution) where the proportion of DM to DE is constant, i.e., $\xi =\xi_0$. The DM, DE, and total (DM+DE), energy densities can be written as functions of the scale factor as
\ba{0.7}
	\epsdm &=& \epsdmnow\left(\frac{a_0}{a}\right)^{3\left(1 +w^* \right)}
	\,,
	\\
	\label{0.7a}\epsde &=& \epsdenow\left(\frac{a_0}{a}\right)^{3\left(1 +w^* \right)}
	\,,
	\\
	\label{0.7b}\bar\varepsilon_{\textrm{tot}} &=& \left(\epsdmnow+\epsdenow\right) \left(\frac{a_0}{a}\right)^{3\left(1 +w^* \right)}
	\,,
\ea
where $w^* \equiv - \gamma/(1+\xi_0)=w/(1+\xi_0)$. Thus $\bar\varepsilon_{\textrm{tot}}$ behaves as a perfect fluid with a constant parameter of EoS $w^*$, a case which was analysed within the Mechanical Approach in \cite{BUZ1} (however as we will see below the results are not the same at the perturbative level). In the case of $w^*=-1$, we find that $\bar\varepsilon_{\textrm{tot}}$ (as well as its constituent parts) behaves as a cosmological constant. In order to obtain late-time acceleration we must impose $w^*<-(1/3)$.

Due to the fact that the ratio $\xi=\xi_0=\const$, it seems that there is no energy density transfer between DM and DE. However, this is not so. To show it, let us consider the case $\gamma <0$, which implies $Q<0$. In this case $w, w^{*}>0$, therefore,
\be{0.7c}
\gamma <0 \quad \Rightarrow \quad 1+w^*>1\,,\quad 1+w^*=1+\frac{w}{1+\xi_0}<1+w
\,,
\ee
and the comparison of Eqs.~\rf{2.11a}, \rf{0.7}, and \rf{0.7a} demonstrates that DM (DE) decreases with the growth of the scale factor faster (slower) in the presence of interaction. Obviously, this happens due to the energy density transfer from DM to DE which ensures the constancy of the ratio $\xi =\xi_0$. Similarly, we can show that in the case $\gamma >0$, we obtain $Q>0$, and there is an energy density transfer from DE to DM.
\\
\subsection{$w+\gamma<0$}
\
	
In this case DE becomes dominant at late-time, since $\xi(a\rightarrow0)=+\infty$ and $\xi(a\rightarrow+\infty)=0$. Therefore the solutions of equations~\eqref{0.5} and \eqref{0.6} have the following limiting behaviours for DM
\ba{0.8}
	\epsdm(\xi\gg1) \simeq \left(\frac{a_0}{a}\right)^{3}
	\quad
	\textrm{and}
	\quad
	\epsdm(\xi\ll1) \simeq \left(\frac{a_0}{a}\right)^{3(1-\gamma)}
	\,,
\ea
and for DE
\ba{0.9}
	\epsde(\xi\gg1) \simeq \left(\frac{a_0}{a}\right)^{3(1+w+\gamma)}
	\quad
	\textrm{and}
	\quad
	\epsde(\xi\ll1) \simeq \left(\frac{a_0}{a}\right)^{3(1+w)}
	\,.
\ea
We see that at very late-time the interaction affects the evolution of DM, which behaves as a perfect fluid with an effective parameter of state $-\gamma$, while essentially leaving DE unaffected.
If DE is to be responsible for the current acceleration of the Universe, then $w<-1/3$.
	
	The behaviour in \eqref{0.8} and \eqref{0.9} is precisely what is observed in the formal solutions of equations~\eqref{0.5} and \eqref{0.6} (see \cite{ABZ} for a derivation of these solutions)
\ba{0.10}
	\epsdm &=& \epsdmnow\left(\frac{a_0}{a}\right)^{3(1-\gamma)}
	\left[\frac{1+\xi_0\left(a/a_0\right)^{3(w+\gamma)}}{1+\xi_0}\right]^{-\frac{\gamma}{w+\gamma}}
	\nn\\
	&=& \epsdmnow\left(\frac{\xi_0}{\xi}\right)^{\frac{1-\gamma}{w+\gamma}}
	\left[\frac{1+\xi}{1+\xi_0}\right]^{-\frac{\gamma}{w+\gamma}}\,,
\ea
\ba{0.11}
	\epsde &=& \epsdenow\left(\frac{a_0}{a}\right)^{3(1+w)}
	\left[\frac{1+\xi_0\left(a/a_0\right)^{3(w+\gamma)}}{1+\xi_0}\right]^{-\frac{\gamma}{w+\gamma}}
	\nn\\
	&=& \epsdenow\left(\frac{\xi_0}{\xi}\right)^{\frac{1+w}{w+\gamma}}
	\left[\frac{1+\xi}{1+\xi_0}\right]^{-\frac{\gamma}{w+\gamma}}\,.
\ea
Eqs.~\rf{0.8} - \rf{0.11} show that, in spite of the fact that DE dominates at the late stage of the Universe evolution, the energy density transfer exists in both directions depending on the sign of $\gamma$. Indeed, for positive $\gamma >0$, the energy density of DM decays slower than $1/a^3$ due to the energy density transfer from DE to DM and vise versa in the case of $\gamma <0$.

Notice that since $\xi$ goes to zero in the future, as $a\rightarrow+\infty$, we can write the following Maclaurin series
\ba{0.12}
	\left[\frac{1+\xi}{1+\xi_0}\right]^{-\frac{\gamma}{w+\gamma}}
	= \left(\frac{1}{1+\xi_0}\right)^{-\frac{\gamma}{w+\gamma}}
	\sum_{i=0}^{+\infty}\alpha_i\xi^i\,.
\ea
From the definition of the coefficients of a Maclaurin series, we obtain the expression
\ba{0.12a}
	\alpha_i &\equiv& \frac{1}{i!}\left.\left[\frac{d^i }{{d\xi}^i}\left(1+\xi\right)^{-\frac{\gamma}{w+\gamma}}\right]\right|_{\xi =0}
	\nn\\
	&=&\frac{1}{i!}
		\left[-\frac{\gamma}{w+\gamma}\right]
		\left[-\frac{\gamma}{w+\gamma}-1\right]...
		\left[-\frac{\gamma}{w+\gamma}-(i-1)\right]
	\nn\\
	&=&\frac{1}{i!}\left(\frac{-1}{w+\gamma}\right)^{i}
		\gamma
		\left[\gamma+\left(w+\gamma\right)\right]...
		\left[\gamma+(i-1)\left(w+\gamma\right)\right]
		\,.
\ea
Therefore, the coefficients $\alpha_i$ are given by
\ba{0.13}
	\alpha_0 &=& 1\,,
	\\	
	\label{0.13a}\alpha_{i>0} &=& \frac{(-1)^i}{i!}\prod_{j=1}^i\frac{\gamma + (j-1)(w+\gamma)}{w+\gamma}
	= \frac{(-1)^i}{i!}\left(\frac{\gamma}{w+\gamma}\right)_{i}\,.
\ea
In Eq.~\eqref{0.13a}, we have introduced the Pochhammer symbol, $(x)_i$, to designate the ascending factorial $(x)_i=x(x+1)(x+2)...(x+i-1)$ \cite{Abramowitz,NIST}. We can now write the energy density of DM and DE as
\ba{0.14}
	\epsdm &=& \epsdenow \left(\frac{1}{1+\xi_0}\right)^{\frac{\gamma}{|w+\gamma|}} \left(\frac{a_0}{a}\right)^{3(1+w)}
		\sum _{i=0}^{+\infty}\alpha_i \xi^{i+1}\,,
		\\
	\label{0.15}
	\epsde &=& \epsdenow \left(\frac{1}{1+\xi_0}\right)^{\frac{\gamma}{|w+\gamma|}} \left(\frac{a_0}{a}\right)^{3(1+w)}
		\sum _{i=0}^{+\infty}\alpha_i \xi^{i}\,,
\ea
where we took into account that $\epsdenow = \epsdmnow/\xi_0$.

\


\section{Cosmological perturbations in the late Universe: mechanical approach}

\label{Sec3}

\setcounter{equation}{0}

\

As we have already mentioned in the introduction, we consider the Universe at its late stage of evolution and deep inside the cell of uniformity. At such scales the Universe is highly inhomogeneous. The inhomogeneities in the form of galaxies, groups and clusters of galaxies have already formed. To describe the dynamical behaviour of the inhomogeneities, the discrete cosmology (mechanical) approach was proposed in \cite{EZcosm1,EZcosm2,EKZ2} (see also the related methods developed in \cite{Villa,Uemh,Chis,Adamek,Ellis}). The above mentioned inhomogeneities affect the background model, that is, the background metric as well as the background matter/energy in the model. The scalar perturbations of a Friedmann-Lema\^itre-Robertson-Walker metric in the Newtonian gauge read:
\be{3.1}
ds^2\approx a^2\left[(1+2\Phi)d\eta^2-(1-2\Phi)\gamma_{\alpha\beta}dx^{\alpha}dx^{\beta}\right]\,.
\ee
In the late Universe, the inhomogeneities (e.g. galaxies and groups of galaxies) have non-relativistic peculiar velocities. In the mechanical approach \cite{EZcosm1,EZcosm2}, we also consider a special form of matter sources (playing, e.g., the role of DM and DE) which are coupled to these inhomogeneities \cite{scal field,coupled}. By this we mean that the fluctuations of the energy density of such matter sources are concentrated around the inhomogeneities. Therefore, the peculiar velocities of these fluctuations are also non-relativistic, and in the system of equations for the scalar perturbations we can drop the terms containing the peculiar velocities as compared with their energy density and pressure fluctuations.
Then, the gravitational potential $\Phi$ takes the form \cite{EZcosm1,EZcosm2,BUZ1,Laslo2,CPL,Chaplygin} :
\be{3.1a}
\Phi({\bf r},\eta)=\frac{\varphi({\bf r})}{c^2 a}\,,
\ee
where
the function $\varphi$ (the "comoving" gravitational potential) depends only on the comoving spatial coordinates%
\footnote{\label{new}{It is worth noting that in general the radius-vectors $\mathbf{r}_i$ of the inhomogeneities (see e.g. Eq. \rf{4.4} below) depend on time.
However, the peculiar velocities are so small (highly non-relativistic at present) that effectively at each moment in time we can calculate the potential just by knowing the position of the inhomogeneities. In this sense the mechanical approach is like a quasi-static approximation, i.e., the particles move so slowly that, at each moment in time, we can compute all relevant quantities just as if they were at rest. As it was shown in \cite{EZcosm2}, the peculiar velocities of the inhomogeneities behaves as $v\sim 1/a$. Hence, the larger is the scale factor, $a$, i.e. at a later time of the evolution of the Universe, the better this approach works. It is as well worth noting that this quasi-static approximation is considered with respect to the comoving space. Therefore, the scale factor of the Universe still has a dynamical behaviour. Summarizing, in the mechanical approach, the gravitational potentials of the inhomogeneities are defined by their positions and not through their velocities. This approximation is similar to the one employed in astrophysics (see section 106 in  \cite{Landau}). The main difference is that here we take into account the expansion of the Universe (and we additionally consider other perfect fluids in the coupled form). After determining the gravitational potential of the system, we can use it to describe
the relative motion of the inhomogeneities (e.g. galaxies) \cite{EKZ2}.}}
${\bf r}$. The gravitational potential $\Phi$ satisfies the following system of linearised Einstein equations:
\ba{3.2}
&{}&\Delta\Phi+3\mathcal{K}\Phi=
\frac{1}{2}\kappa a^2\left(\delta T^0_{0}+\delta\varepsilon_{\mathrm{rad}}+\delta\varepsilon_{\mathrm{DM}}+\delta\varepsilon_{\mathrm{DE}}\right)\,,\\
\label{3.3} &{}&\left(\mathcal{H}'-\mathcal{H}^2-\mathcal{K}\right)\Phi=
\frac{1}{2}\kappa a^2\left(\delta p_{\mathrm{DE}}+\delta p_{\mathrm{rad}}\right)\,.
\ea
Here, $\Delta$ is the Laplace operator defined with respect to the spatial metric $\gamma_{\alpha\beta}$, and the fluctuation of the energy density of the non-interacting non-relativistic matter (at the linear approximation with respect to $\Phi$) reads \cite{EZcosm1}
\be{3.4}
\delta T^0_{0}=\frac{\delta\rho_{\mathrm{c}}\, c^2}{a^3}+\frac{3\overline{\rho}_{\mathrm{c}}\, c^2\Phi}{a^3}
= \frac{\delta\rho_{\mathrm{c}}\, c^2}{a^3}+\frac{3\overline{\rho}_{\mathrm{c}}\,\varphi}{a^4}\,,
\ee
where the fluctuation of the comoving rest mass density is given as $\delta\rho_{\mathrm{c}}=\rho_{\mathrm{c}}-\overline\rho_{\mathrm{c}}$.

With the help of \rf{2.4}, Eq.~\rf{3.3} reads (make mention to how $\overline T^0_0$ was replaced)
\be{3.5}
\left[\frac{\overline\rho_{\mathrm{c}}\, c^2}{a^3}+
\frac43\overline\varepsilon_{\mathrm{rad}} +\overline \varepsilon_{\mathrm{DM}} +(1+w)\overline \varepsilon_{\mathrm{DE}}
\right]\frac{\varphi}{c^2a}=-w\delta \varepsilon_{\mathrm{DE}}-\frac13 \delta\varepsilon_{\mathrm{rad}}\,,
\ee
where we also took into account that $\delta p_{\mathrm{rad}} = (1/3) \delta\varepsilon_{\mathrm{rad}}$. From this EoS and within the mechanical approach, where the peculiar velocities are assumed to be non-relativistic, we can easily get that $\delta\varepsilon_{\mathrm{rad}} \sim 1/a^4$ (we refer the reader to the appendix of \cite{CPL} for a derivation of this result).
The important point is that, according to Eq.~\rf{3.4}, the accuracy of our approach is $O(1/a^4)$ \footnote{See e.g. the footnote 2 in \cite{CPL} where we justified this approximation.}. This means that in our perturbed equations, in the late Universe we should drop all the terms which behave as $1/a^n$, with $n>4$. For example, in the above equation \rf{3.5} the term $\overline\varepsilon_{\mathrm{rad}}(\varphi/a)\sim 1/a^5$ should be omitted. Following this reasoning, we get for the fluctuations of interacting dark energy
\be{3.6}
-w\delta \varepsilon_{\mathrm{DE}}=
\left[\frac{\overline\rho_{\mathrm{c}}\, c^2}{a^3}+
\overline \varepsilon_{\mathrm{DM}} +(1+w)\overline \varepsilon_{\mathrm{DE}}
\right]\frac{\varphi}{c^2a}+\frac13 \delta\varepsilon_{\mathrm{rad}}\,.
\ee
We can easily verify that the same expression for $\delta \varepsilon_{\mathrm{DE}}$ follows from the total conservation equation \cite{CPL}:
\be{3.7}
\sum_{i}\left[\delta p_{\mathrm{X}_i} + \left(\bar \varepsilon_{\mathrm{X}_i}+\bar p_{\mathrm{X}_i}\right)\Phi\right]=0\,,
\ee
where the summation should be taken over all matter/energy constituents of the model. Keeping in mind that the contribution of the $\Lambda$-term disappears from this equation, we must consider in our case the non-interacting non-relativistic matter and radiation, and the interacting dark sector components.

Next, it is convenient to split $\delta\varepsilon_{\mathrm{rad}}$ into two parts:
\be{3.8}
\delta\varepsilon_{\mathrm{rad}}=\delta\varepsilon_{\mathrm{rad1}}+\delta\varepsilon_{\mathrm{rad2}}\,,
\ee
where
\be{3.9}
\delta\varepsilon_{\mathrm{rad1}} \equiv -\frac{3\overline\rho_{\mathrm{c}}\varphi}{a^4}\,,
\ee
and $\delta\varepsilon_{\mathrm{rad2}}$ accounts for a possible contribution of DE to the radiation fluctuations \cite{CPL}.
In the $\Lambda$CDM model, the fluctuation of radiation exactly coincides with the expression \rf{3.9} \cite{EZcosm2}. Then, Eq.~\rf{3.6} takes the form
\be{1.1}
	\delepsde =
		-\frac{1}{3w} {\delepsrad}_2
	-\frac{1}{w}\left[
		\epsdm
		+(1+w)\epsde
	\right]\frac{\varphi}{c^2a}\,.
\ee

If we differentiate equation \eqref{1.1}, we obtain
\be{1.2}
	\delepsde' =
	\frac{4\cH}{3w} {\delepsrad}_2
	+ \frac{\cH}{w}\left[
		\epsdm
		+(1+w)\epsde
	\right]\frac{\varphi}{c^2a}
	-\frac{1}{w}\bar\varepsilon_{\textrm{tot}}'\frac{\varphi}{c^2a}
	-\epsde'\frac{\varphi}{c^2a}
	\,.
\ee
Substituting the conservation equations \rf{2.8} and \rf{2.10} in Eq.~\eqref{1.2}, we find after some algebra
\be{1.3}
	\delepsde'
	= -4\cH\delepsde
	+3\cH(1+w)\epsde\frac{\varphi}{c^2a}
	+Q\frac{\varphi}{c^2a}
	\,.
\ee
Notice that if we turn off the interaction term, we obtain the following solution to $\delepsde$
\ba{1.3a}
	\delepsde = \frac{c_1}{a^4} - \frac{1+w}{c^2w}\epsdenow\frac{a_0^{3+3w}}{a^{4+3w}}\varphi
	\,,
\ea
where $c_1$ is an integration constant. This result is in accordance with the analysis done in \cite{BUZ1}.
On the other hand, the perturbed conservation equation for $\delepsde$ reads (see e.g. the Appendix in \cite{CPL})
\be{1.4}
	\delepsde' + 3\cH\left(1+w\right) \delepsde + \left(1+w\right) \epsde(-3\Phi') = -\dQ\,.
\ee
When we turn off the interaction term the solution of this equation is
\ba{1.4a}
	\delepsde = \frac{c_2}{a^{3+3w}} -\frac{1+w}{c^2(-1/3)}\epsdenow\frac{a_0^{3+3w}}{a^{4+3w}}\varphi
	\,,
\ea
where $c_2$ is an integration constant. Notice that Eqs.~\eqref{1.3a} and \eqref{1.4a} are consistent only in the cases $w=-1/3,-1$ with $c_1=c_2=0$. These are the physically acceptable values of $w$ found in \cite{BUZ1}.\footnote{The case $w=1/3$ and $\epsdenow=0$ is also mathematically consistent but this is just the case of a Universe filled by non-interacting dust and radiation - unsuitable to describe the current Universe. Indeed it is of no interest to the interacting model we are analysing.}

Using Eqs.~\rf{1.1} and \eqref{1.3}, we obtain from \rf{1.4}:
\be{1.5}
	\dQ =
		\cH\left[
		\frac{3w-1}{w}\epsdm
		-\frac{(3w+1)(1+w)}{w}\epsde
	\right]\frac{\varphi}{c^2a}
	+\cH\frac{3w-1}{3w} {\delepsrad}_2
	-Q\frac{\varphi}{c^2a}
	\,.
\ee
The results obtained so far in this section are independent of the form of $Q$ and $\dQ$. Let us now consider the exact form of $Q$, as given by \rf{2.11}.
Perturbing this interaction term, we find
\be{1.6}
	\dQ = 3\cH\gamma\left[\left(\frac{1}{1+\xi}\right)^2\delepsdm + \left(\frac{\xi}{1+\xi}\right)^2\delepsde\right]\,.
\ee
Notice that, while the expression in \rf{2.11} is not obtained from a covariant formulation of the interaction, which means that there is no {\em a priori} prescription on how to obtain the perturbation $\dQ$, in our approach the implementation of the perturbed interaction term derived in Eq.~\rf{1.6} was done consistently with the constraints derived from the perturbed conservation equations for the total matter.

Equation \rf{1.6} enables us to get $\delepsdm$
\ba{1.7}
	\delepsdm
	&=&
	\left[
		\frac{3w-1}{3w\gamma}\left(1+\xi\right)^2
		+ \frac{\xi^2}{w}
	\right]\frac{1}{3}{\delepsrad}_2
	\nn\\&&
	+\left\{
		\frac{(3w-1)\xi-(3w+1)(1+w)}{3w\gamma}\left(1+\xi\right)^2
		+ \frac{(1+\xi)\xi^2}{w}
		-\xi
	\right\}\epsde\frac{\varphi}{c^2a}\,.
\ea
Here we have used \eqref{1.1} and \eqref{1.5} to eliminate $\dQ$ and $\delepsde$.

We are now ready to go back to the $(00)$ component of the perturbed Einstein equations \rf{3.3}, which we can re-write as
\be{1.8}
	\Delta\varphi+3\cK\varphi - \frac{\kappa c^4}{2}\delta\rho_{\mathrm{c}} =
	\frac{\kappa c^2a^3}{2}
	\left(
		{\delepsrad}_2
		+\delepsdm
		+\delepsde
	\right)\,.
\ee
By substituting \eqref{1.1} and \eqref{1.7} in this equation, we find
\ba{1.9}
	&{}&\Delta\varphi+3\cK\varphi - \frac{\kappa c^4}{2}\delta\rho_{\mathrm{c}} =
		\left[1 - \frac{1-\xi^2}{3w} + \frac{3w-1}{9w\gamma}\left(1+\xi\right)^2\right]
	\frac{\kappa c^2a^3}{2}{\delepsrad}_2
	\nn\\
	&{}&-\left[
		1
		+\frac{1-\xi^2}{w}
		-\frac{(3w-1)\xi-(3w+1)(1+w)}{3w\gamma}\left(1+\xi\right)
	\right](1+\xi)\epsde\frac{\kappa c^2a^2\varphi}{2}\,.
\ea
The left-hand-side (l.h.s.) of this equation does not depend on time. Therefore, the same should be true within the adopted accuracy for the right-hand-side (r.h.s.) of Eq.~\eqref{1.9}. In general, there are two possibilities to achieve this: either the terms in r.h.s. compensate each other, or each of them do not depend on time separately. Some terms may vanish because according to our accuracy, we define fluctuations of the energy densities and pressure up to terms $O(1/a^4)$, inclusively. Bearing in mind that r.h.s. of \rf{1.8} has been multiplied by $a^3$, on r.h.s. of \rf{1.9} we must drop all terms $1/a^n$, with $n>1$. Now, we should analyse under which conditions r.h.s. of \rf{1.9} is self-consistent. It makes sense to consider separately the cases of $w+\gamma=0$ and $w+\gamma<0$.

Before continuing we compare the results obtained so far, in particular Eq.~\eqref{1.9}, with those obtained in the mechanical approach for a non-interacting DE fluid with constant EoS parameter $w$ \cite{CPL}. Intuitively, to recover the results of \cite{CPL} we should set $\gamma=0$ as this switches off the interaction both at the background, $Q$, and perturbative, $\delta Q$, levels. However, in Eq.~\eqref{1.9} we find terms inversely proportional to $\gamma$ that diverge as we take the limit $\gamma\rightarrow0$. This comes from the fact that Eq.~\eqref{1.7}, as derived from Eq.~\eqref{1.6}, is not defined when $\gamma=0$, i.e., in the absence of interaction we cannot derive from Eq.~\eqref{1.7} a relation between the perturbations of DM and DE, and consequently between $\delepsdm$, ${\delepsrad}_2$, and $\varphi$. A way around this problem is to notice that in the absence of interaction, the interacting DM component should vanish, since all non-interacting non-relativistic matter is already included in the matter distribution $\bar\rho_c$. Therefore, when we switch off the interaction we should set to zero $\epsdm$, $\delepsdm$, and $\xi$. Consequently in Eq.~\eqref{1.9} we should drop all the terms coming from the term $\delepsdm$ in Eq.~\eqref{1.8}. To better identify such terms, let us set $\delepsdm=0$ and $\xi=0$ in Eq.~\eqref{1.7}. The two remaining terms of r.h.s. of the equation are those that we should drop from Eq.~\eqref{1.9} after setting $\xi=$ in that equation. After the comparison with the equation obtained in \cite{CPL} we find that we do recover the result for the case of no interaction.

\subsection{$w+\gamma=0$}
\label{sec3.1}
\

Let us first analyse the case of $w+\gamma=0$, which corresponds to a self-similar solution of DM and DE. As we have seen before, in this case we have $\xi =\xi_0$, $w=(1+\xi_0)w^*$, $\gamma=-(1+\xi_0)w^*$, and $\epsdm$ and $\epsde$ are described by Eqs.~\rf{0.7} and \rf{0.7a}, respectively.
If we substitute these expressions in Eq.~\eqref{1.9} we find that r.h.s. of this equation has the structure
\be{1.11}
	r.h.s
	= A_1 \frac{a_0}{a} + A_2 \left(\frac{a_0}{a}\right)^{1+3w^*}
	\,,
\ee
where
\be{1.11aa}
	A_1\equiv\frac{\left(3w^*-1\right)^2}{9(w^*)^2}
	\frac{\kappa c^2a_0^3}{2}{\delepsrad}_{2,0}
	\,,
\ee
and
\be{1.11aaa}
	\qquad
	A_2\equiv\frac{
		1
		+w^*
		+3\xi_0(w^*)^2
	}{3(w^*)^2}\left(1+\xi_0\right)\epsdenow\frac{\kappa a_0^2\varphi}{2}\,.
\ee
Here we took into account the dependence of the fluctuation of radiation on the scale factor:
\be{1.11aaaa}
{\delepsrad}_{2}({\bf r},\eta) = {\delepsrad}_{2,0}({\bf r})\left(\frac{a_0}{a}\right)^4\,,
\ee
where ${\delepsrad}_{2,0}({\bf r})$ is the distribution of the fluctuations at the present time.
The two terms on r.h.s. of equation \eqref{1.11} cannot compensate each other as they have different dependencies on $a$: the first term goes as $(a_0/a)$ while the second goes as $(a_0/a)^{1+3w^*}$ (we are assuming that $w^*\neq0$ as $w^*=0$ implies $w=0$). Therefore, in order for r.h.s. to be time independent we need to impose, for each term separately, that it either vanishes or is independent of time.

Regarding the first term we have to impose that $A_1=0$ which results in either ${\delepsrad}_{2,0}=0$, or in $w^*=1/3$, i.e., $\epsdm$ and $\epsde$ (and, consequently, $\bar\varepsilon_{\textrm{tot}}$) behave as radiation.

Regarding the second term we have three possibilities. The first is that $w^*>0$, in which case the term can be dropped altogether within the accuracy of the mechanical approach. As a consequence, we find that interacting DM and DE do not contribute to the gravitational potential. The second possibility is that $w^*=-1/3$, in which case the term is independent of time and the late Universe is neither accelerating nor decelerating\footnote{The acceleration of the Universe takes place if the following combination is positive: $-\epsdm -(1+3w)\epsde = -(1+\xi_0)(1+3w^{*})\epsde >0$. Obviously, this happens for $w^{*}<-1/3$ (we remind that both $\xi_0$ and $\epsdenow$ are positive).}.
Finally, the third possibility is that $A_2=0$, which leads to
\be{1.12}
	1+w^* + 3\xi_0(w^*)^2=0
	\quad
	\Rightarrow
	\quad
	\xi_0 = -\frac{1+w^*}{3{w^*}^2}
	\,.
\ee
Notice that since $\xi_0$ is positive by definition, if the previous equation holds we must have $w^*<-1$, i.e., phantom behaviour of dark energy. For a fixed value of $\xi_0$ we get
\be{1.12a}
w^*=\frac{-1\pm \sqrt{1-12\xi_0}}{6\xi_0}\,,\quad 0<\xi_0\leq \frac{1}{12}\,.
\ee
It can be easily seen that $w^* \to -1$ for $\xi_0\to 0$ if we choose the upper sign in \rf{1.12a}. The case of $w^*=-1$ implies that $\xi_0=0$, i.e. $\epsdm=0$, and we arrive to the case of a cosmological constant with no interaction.

\

Finally, we list the acceptable cases found for $w+\gamma=0$:
\begin{enumerate}[(i)]
	\item $w^*>0$, $w^*\neq1/3$ and ${\delepsrad}_2=0$;
	\item $w^*=1/3$ and ${\delepsrad}_2$ may be non-zero;
	\item $w^*=-1/3$ and ${\delepsrad}_2=0$;
	\item $w^*=-1$, $\xi_0=0$, and ${\delepsrad}_2=0$. This case reduces to the cosmological constant with no interaction\footnote{This reduces to $\Lambda$CDM model in absence of interactions on the dark sector. Given that this case has been previously analysed in \cite{EZcosm1}, we will disregard its analysis. The same criterium will be applied whenever we recover $w$CDM models (with $w$ constant) in absence of interactions on the dark sector, which were studied in \cite{BUZ1} (see also the appendix in \cite{CPL}).};
	\item $w^*<-1$ and ${\delepsrad}_2=0$. This case leads to phantom behaviour and $\xi_0$ is defined by the relation \eqref{1.12}.

\end{enumerate}

\subsection{$w+\gamma<0$}
\label{sec3.2}
\

In this case, the expressions for $\epsdm$ and $\epsde$ are given by \eqref{0.15}. Substituting them on r.h.s. of Eq.~\eqref{1.9}, we obtain
\ba{1.14}
	r.h.s
	&=&
\frac{\kappa c^2a_0^3}{18w\gamma}{\delepsrad}_{2,0} \left(\sum_{i=0}^{2}B_i\xi_0^i
	\left(\frac{a_0}{a}\right)^{1+3i|w+\gamma|} \right)\nn \\
	&+& \left(\frac{1}{1+\xi_0}\right)^{-\frac{\gamma}{w+\gamma}} \frac{\kappa a_0^2 \epsdenow }{6w\gamma}
	\left(\sum_{i=0}^{+\infty}D_i\xi_0^{i} \left(\frac{a_0}{a}\right)^{1+3w+3i|w+\gamma|}\right)\varphi
	\,.
\ea
Here, we have introduced the coefficients
\ba{1.14b}
	B_0 &\equiv& (3\gamma +1)(3w-1)\,,
	\\
	B_1 &\equiv& 2(3w-1)\,,
	\\
	B_2 &\equiv& 3(w+\gamma)-1\,,
\ea
and
\ba{1.14d}
	D_i &\equiv& \sum_{j=0}^{i}C_j\alpha_{i-j}\, \; ,
\ea
where
\ba{1.14c}
	C_0 &\equiv& -\left[3(w+\gamma)+1\right](1+w)\,,
	\\
	C_1 &\equiv& -3(1+w)(1+\gamma) - 2w(1+3w)\,,
	\\
	C_2 &\equiv& -3 + 3(w+\gamma) - w(1+3w)\,,
	\\
	C_3 &\equiv& -1 + 3(w+\gamma)\,,
 \\
 C_j &\equiv& 0\,, \quad j\geq 4,
\ea
and the coefficients $\alpha_i$ are defined in Eqs.~\eqref{0.13} and \rf{0.13a}.

Before continuing, let us investigate which terms we can drop from Eq.~\eqref{1.14} within the accuracy of the mechanical approach.

With respect to the three terms that are related to the coefficients $B_i$, we find that we need only to keep the term proportional to $B_0$, which goes like $a_0/a$, as this is already in the boundary of the accuracy. The other two terms are outside the accuracy of the approach and can be dropped.

For the terms that are related to the coefficients $D_i$, we find that the maximum order that needs to be considered is defined by the biggest integer $i_\textrm{max}^D$ that satisfies the inequality
\ba{14.a}
	&&1+3w+3i_\textrm{max}^D|w+\gamma|\leq1
	\,,
	\quad
	\Rightarrow
	\quad
	i_\textrm{max}^D = E\left(-\frac{w}{|w+\gamma|}\right)
	\,,
\ea
where $E(x)$ is the floor function that returns the largest integer smaller than $x$ \cite{NIST}. The value of $i_\textrm{max}^D$ depends then on the ratio $\gamma/w$.
Taking into account the previous considerations we can reduce \eqref{1.14} to
\ba{1.15}
	r.h.s &=& \frac{\kappa c^2a_0^3}{18w\gamma}{\delepsrad}_{2,0}B_0\left(\frac{a_0}{a}\right)\nn \\
&+&\left(\frac{1}{1+\xi_0}\right)^{-\frac{\gamma}{w+\gamma}} \frac{\kappa a_0^2 \epsdenow }{6w\gamma}
	\left(\sum_{i=0}^{i_\textrm{max}^D}D_i\xi_0^{i} \left(\frac{a_0}{a}\right)^{1+3w+3i|w+\gamma|}\right)\varphi
	\,.
\ea
Within the region $w+\gamma<0$ of the parameter space, we have three distinct scenarios:


\subsubsection{Energy transfer from DM to DE I: $w>0\; \to \; \gamma <0$}

\

In the case of $w>0$, we find from equation \eqref{14.a} that $i_\textrm{max}^D <0$. Therefore, all the terms coming from the $D_i$'s can be neglected under the accuracy of the mechanical approach. Therefore, the equation \eqref{1.15} reduces to
\ba{1.16}
	r.h.s = \frac{\kappa c^2a_0^3}{18w\gamma}{\delepsrad}_{2,0}B_0\left(\frac{a_0}{a}\right)
	\,,
\ea
and, in order for it to be independent of time, we must impose either that $ {\delepsrad}_2=0$, i.e. the interacting DE and DM components do not have a contribution to the perturbation of radiation, or that $B_0=0$, i.e.
\ba{1.17}
	B_0 = (3w-1)(3\gamma+1) = 0
	\,.
\ea
For positive values of $w$ we find that the only allowed roots of the previous equation are $w=1/3$, and $\gamma=-1/3$ (with $0<w<1/3$).

Therefore, we obtain three possible cases:
\begin{enumerate}[(i)]
	\item ${\delepsrad}_2=0$, $w>0$, $\gamma<-w, \gamma \neq -1/3$;
	\item $w=1/3$, $\gamma<-1/3$, and ${\delepsrad}_2$ may be non-zero;
	\item $\gamma=-1/3$, $0<w<1/3$, and ${\delepsrad}_2$ may be non-zero.
\end{enumerate}

\
\subsubsection{Energy transfer from DM to DE II: $w<0$ and $\gamma<0$}
\

In this case, we have $|w+\gamma|>-w$ and we find that $i_\textrm{max}^D=0$, i.e., only the term proportional to $D_0$ in Eq.~\eqref{1.15} needs to be kept:
\ba{1.18}
	r.h.s = \frac{\kappa c^2a_0^3}{18w\gamma}{\delepsrad}_{2,0} B_0\left(\frac{a_0}{a}\right)
	+ \left(\frac{1}{1+\xi_0}\right)^{-\frac{\gamma}{w+\gamma}} \frac{\kappa a_0^2 \epsdenow }{6w\gamma} D_0\left(\frac{a_0}{a}\right)^{1+3w}\varphi\,.
\ea
Notice that the two terms in this equation correspond to different powers of the scale factor and, consequently, they cannot cancel each other. Therefore, in order for r.h.s. to be independent of time we must analyse these terms individually.

Regarding the term involving $B_0$ we must again impose that either $ {\delepsrad}_2=0$, or that $B_0=0$. For negative $w$ we find that equation \eqref{1.17} only has one root: $\gamma=-1/3$.

Regarding the term proportional to $D_0$ we must either impose $D_0=0$ or $w=-1/3$. The condition $D_0=0$ can be written as
\ba{1.19}
	D_0 = C_0\alpha_0= -(1+w)(1+3w+3\gamma) =0
	\,.
\ea
The roots of this equation are $w=-1$ and $w+\gamma=-1/3$. Regarding the second root, notice that since we have assumed $w<0$ and $\gamma<0$, if $w+\gamma=-1/3$ then $-1/3<w<0$ and $-1/3<\gamma<0$.

Thus, we have the following possibilities that are viable within the mechanical approach:
\begin{enumerate}[(i)]
	\item $-1/3<w<0$, $w+\gamma=-1/3$, and ${\delepsrad}_2=0$;
	\item $w=-1/3$, $\gamma=-1/3$, and ${\delepsrad}_2$ can be non-zero;
	\item $w=-1/3$, $\gamma<0$, $\gamma\neq-1/3$, and ${\delepsrad}_2=0$;
	\item $w=-1$, $\gamma=-1/3$, and ${\delepsrad}_2$ can be non-zero;
	\item $w=-1$, $\gamma<0$, $\gamma\neq-1/3$, and ${\delepsrad}_2=0$.
\end{enumerate}

\
\subsubsection{Energy transfer from DE to DM: $w<0$ and $0<\gamma<-w$}
\

In this case $i_\textrm{max}^D\geq1$, therefore we have to consider at least the two terms with coefficients $D_0$ and $D_1$.
Then, Eq.~\rf{1.15} reads
\ba{1.20}
	r.h.s &=& \frac{\kappa c^2a_0^3}{18w\gamma}{\delepsrad}_{2,0} B_0\left(\frac{a_0}{a}\right)\nn \\
	&+& \left(\frac{1}{1+\xi_0}\right)^{-\frac{\gamma}{w+\gamma}} \frac{\kappa a_0^2 \epsdenow }{6w\gamma}
	\sum_{i=0}^{i_\textrm{max}^{D}\geq1}D_i\xi_0^{i} \left(\frac{a_0}{a}\right)^{1+3w+3i|w+\gamma|}\varphi
	\,.
\ea

Let us consider first the possibility that some of the terms involving $D_i$ can cancel the term with $B_0$. This may happen if there is some integer $i_0^D\leq i_\textrm{max}^D$ satisfying the condition
\ba{1.21}
	1+3w+ 3i_0^D|w+\gamma|=1 \Rightarrow i_0^D = -\frac{w}{|w+\gamma|}\,.
\ea
Comparing equations \eqref{14.a} and \eqref{1.21}, we find that, if there exists $i_0^D$ such that equation \eqref{1.21} is satisfied, then $i_0^D=i_\textrm{max}^D$ and
\ba{1.21a}
	\gamma=\frac{i_0^D-1}{i_0^D}|w|
	\,.
\ea
Notice that $i_0^D=1$, would imply $\gamma=0$, i.e., no interaction, while $i_0^D=0$ would imply $w=0$, two cases which are not of interest. Since we have excluded the parameters $\gamma,w = 0$, then $i_0^D\geq 2$. Therefore, the terms in $D_0$ and $D_1$ never couple with the term in $B_0$ and both must vanish except, if for some positive integer $i_1^D\leq i_\textrm{max}^D$, we have
\ba{1.22}
	1+3w+ 3i_1^D|w+\gamma|=0 \Rightarrow i_1^D = \frac{1}{|w+\gamma|}\left( -\frac{1}{3}-w\right)
	\,.
\ea
If the previous condition is satisfied, the corresponding term has no dependency on the scale factor. Notice that in Eq.~\eqref{1.22} only for $w\leq-1/3$ we do have $i_1^D\geq0$.
Also, if $i_1^D$ exists, then
\ba{1.22a}
	\gamma=\frac{1}{3i_1^D}+\frac{i_1^D-1}{i_1^D}|w|\,.
\ea

Let us now focus our analysis on the coefficients $D_0$ and $D_1$.
The coefficient $D_0$ is given by
\ba{1.23}
	D_0 \equiv C_0\alpha_0= -(1+w)(1+3w+3\gamma)\,.
\ea
Equating this expression to zero gives two roots: $w=-1$ and $w+\gamma=-1/3$. Regarding the second root $w+\gamma=-1/3$, notice that since in this section we assume $w<0$ and $0<\gamma<-w$, we get $w<-1/3$.

The coefficient $D_1$ is given by
\ba{1.24}
	D_1 \equiv C_0\alpha_1 + C_1\alpha_0
= -2\left[(1+w) + w(1+3w) \right] - \frac{w(1+w)}{w+\gamma}\,.
\ea
It can be easily verified that neither $w=-1$ nor $w+\gamma=-1/3$ satisfy the equation $D_1=0$.
In fact, equating r.h.s of Eq.~\eqref{1.24} to zero gives the solution
\ba{1.25}
	\gamma = - w \frac{3(1+w) + 2w(1+3w)}{2(1+w) + 2w(1+3w)} = -w - \frac{w(1+w)}{2(1+w) + 2w(1+3w)}\,.
\ea
This expression for $\gamma$ satisfies the condition $0<\gamma<-w$ only for $w<-1$.

From the previous analysis we find that there is no combination of parameters $w$ and $\gamma$ such that $D_0$ and $D_1$ vanish simultaneously. We are then left with two alternatives: either $D_0=0$ and $i_1^D=1$, or $D_1=0$ and $i_1^D=0$. We will next analyse them.
\begin{enumerate}[(i)]
	\item $D_0=0$ and $i_1^D=1$: from $D_0=0$, we obtain $w=-1$ or $w+\gamma=-1/3$, while from $i_1^D=1$, we obtain $\gamma=1/3$. Thus the allowed values are $w=-1$ with $\gamma=1/3$, or $w=-2/3$ with $\gamma=1/3$.
	We now take a closer look at the two allowed cases in this item:
\begin{enumerate}
\item For $w=-1$ with $\gamma=1/3$ we have that $i_\textrm{max}^D=1$ (see Eq.~\rf{14.a}) and therefore we do not have to worry about higher order terms as they are outside the accuracy of the mechanical approach. In this case we find from equation \eqref{1.14b} that $B_0=-8$. This means that we must impose that ${\delepsrad}_{2,0} =0$ for r.h.s. of equation \eqref{1.20} to be independent of time.

\item However, for $w=-2/3$ with $\gamma=1/3$, we have $i_\textrm{max}^D=2$, and the condition \eqref{1.21} is satisfied, meaning that the terms in $B_0$ and $D_2$ are coupled and the sum of these two terms must vanish. Using the definitions of $B_0$ and $D_2$, we obtain:
\ba{1.26}
	B_0 = (3\gamma+1)(3w-1) = -6\,,
\ea
\ba{1.27}
	D_2 &=& C_0\alpha_2 + C_1\alpha_1 + C_2\alpha_0
	\nn\\
		&=& 3\gamma + \frac{-2 + 9w + 3w^2}{2} - \frac{w(3+4w+9w^2)}{2(w+\gamma)} - \frac{w^2(1+w)}{2(w+\gamma)^2}
	\nn\\
		&=& -\frac{22}{3}
	\,.
\ea
As stated above, the sum of the terms in Eq.~\eqref{1.20} that depend on $B_0$ and $D_2$ must vanish. After the substitution of Eqs.~\eqref{1.26} and \eqref{1.27} on such a linear combination and equating it to zero, we obtain:
\ba{1.28}
	3 \frac{\kappa c^2a_0^3}{2}{\delepsrad}_{2,0} \left(\frac{a_0}{a}\right)
	+\frac{11}{2} \frac{\xi_0^2}{1+\xi_0}\kappa a_0^2 \epsdenow \left(\frac{a_0}{a}\right)\varphi=0\,,
\ea
which results in the following relation:
\ba{1.29}
	{\delepsrad}_{2,0} = - \frac{11}{3}\frac{\xi_0^2}{1+\xi_0}\epsdenow\frac{\varphi}{a_0 c^2}\,.
\ea
\end{enumerate}
	
	\item $D_1=0$ and $i_1^D=0$: from $D_1=0$, we obtain $w<-1$ with $\gamma$ given by equation \eqref{1.25}, while $i_1^D=0$ gives $w=-1/3$. Therefore there is no combination of parameters for which this case is allowed within the mechanical approach.
\end{enumerate}


In summary, in this subsection we obtain two further cases that are allowed within the mechanical approach:
\begin{enumerate}[(i)]
	\item ${\delepsrad}_2=0$, $w=-1$, and $\gamma=1/3$;
	\item $w=-2/3$, $\gamma=1/3$, and ${\delepsrad}_2\propto\varphi$, as obtained in equation \eqref{1.29}.
\end{enumerate}

\
\section{Discussion and conclusions}

\label{Sec4}

\setcounter{equation}{0}

\


\begin{table}[t]
\centering
\begin{tabularx}{\textwidth}{ L{.3} C{1.4} C{1.4} C{.9} C{.9} C{.9} C{1.3} C{.9} }
\toprule
	{\bf \# }
	 	& {\bf $w^*\equiv \frac{w}{1+\xi_0}$}	
	 		& {\bf $Q$}
	 			& {\bf $\delta\varepsilon_{\mathrm{rad2}}$}
	 			& { $\delta\varepsilon_\mathrm{DM}$}
	 			& { $\delta\varepsilon_\mathrm{DE}$}
	 				& late-time
 					& { r.h.s. of \rf{1.9}}
\\\midrule
	1.
		& $w^*>0$, $w^*\neq\frac{1}{3}$
			& $Q<0$ (DM$ \rightarrow $DE)
				& 0
				& 0
				& 0
					& deceleration
						& 0
 \\ \hline
	2.
		& $w^*= \frac{1}{3}$
			& $Q<0$ (DM$ \rightarrow $DE)
				& can be non-zero
				& $\propto\delta\varepsilon_{\mathrm{rad2}}$
				& $\propto\delta\varepsilon_{\mathrm{rad2}}$
					& deceleration
						& 0
 \\ \hline
	3.
		& $w^*= -\frac{1}{3}$
			& $Q>0$ (DE$ \rightarrow $DM)
				& $0$
				& $\propto\varphi$
				& $\propto\varphi$
					& zero acceleration
						& const$\times\varphi$
 \\ \hline
	4.
		& $w^*= -1$
				& $Q=0$
				& $0$
				& $0$
				& $0$
					& cosmological constant
						& $0$
 \\ \hline
	5.
		& $w^*< -1$
			& $Q>0$ (DE$ \rightarrow $DM)
				& $0$
				& $\propto\varphi$
				& $\propto\varphi$
					& phantom behaviour
						& $0$
 \\
\bottomrule
\end{tabularx}
\caption{\label{results_table_I} Values of the parameters that are compatible with the mechanical approach for the model with non-linear interaction between DM and DE in the case $w+\gamma=0$, as derived in Subsection \ref{sec3.1}.}
\vspace{.15cm}
\end{table}


\begin{table}[t]
\centering
\begin{tabularx}{\textwidth}{ L{.3} C{1.4} C{1.4} C{.9} C{.9} C{.9} C{1.3} C{.9} }
\toprule
	{\bf \# }
	 	& {\bf $w$}	
	 		& {\bf $\gamma$}
	 			& {\bf $\delta\varepsilon_{\mathrm{rad2}}$}
	 			& { $\delta\varepsilon_\mathrm{DM}$}
	 			& { $\delta\varepsilon_\mathrm{DE}$}
	 				& late-time
 					& { r.h.s. of \rf{1.9}}
\\\midrule
	1.
		& $w>0$, $w\neq\frac{1}{3}$
			& $\gamma<-w<0$, $\gamma\neq-\frac{1}{3}$ (DM$ \rightarrow $DE)
				& 0
				& 0
				& 0
					& deceleration
						& 0
 \\ \hline
	2.
		& $w= \frac{1}{3}$
			& $\gamma<-\frac{1}{3}$ (DM$ \rightarrow $DE)
				& can be non-zero
				& $\propto\delta\varepsilon_{\mathrm{rad2}}$
				& $\propto\delta\varepsilon_{\mathrm{rad2}}$
					& deceleration
						& 0
 \\ \hline
	3.
		& $0<w<\frac{1}{3}$
			& $\gamma=-\frac{1}{3}$ (DM$ \rightarrow $DE)
				& can be non-zero
				& $\propto\delta\varepsilon_{\mathrm{rad2}}$
				& $\propto\delta\varepsilon_{\mathrm{rad2}}$
					& deceleration
						& 0
 \\ \hline
	4.
		& $-\frac{1}{3}<w<0$
				& $\gamma=-\frac{1}{3}-w$ (DM$ \rightarrow $DE)
				& $0$
				& $\propto\varphi$
				& $\propto\varphi$
					& deceleration
						& $0$
 \\ \hline
	5.
		& $w=-\frac{1}{3}$
			& $\gamma<0$, $\gamma\neq-\frac{1}{3}$ (DM$ \rightarrow $DE)
				& $0$
				& 0
				& $\propto\varphi$
					& zero-acceleration
						& const$\times\varphi$
 \\ \hline
	6.
		& $w=-\frac{1}{3}$
			& $\gamma=-\frac{1}{3}$ (DM$ \rightarrow $DE)
				& can be non-zero
				& $\propto\delta\varepsilon_{\mathrm{rad2}}$
				& non-zero
					& zero-acceleration
						& const$\times\varphi$
 \\ \hline
	7.
		& $w=-\frac{2}{3}$
			& $\gamma=\frac{1}{3}$ (DE$ \rightarrow $DM)
				& $\propto\varphi$
				& $\propto\varphi$
				& $\propto\varphi$
					& acceleration
						& const$\times\varphi$
 \\ \hline
	8.
		& $w=-1$
			& $\gamma<0$, $\gamma\neq-\frac{1}{3}$ (DM$ \rightarrow $DE)
				& 0
				& 0
				& 0
					& cosmological constant
						& 0
 \\ \hline
	9.
		& $w=-1$
			& $\gamma=-\frac{1}{3}$ (DM$ \rightarrow $DE)
				& can be non-zero
				& $\propto\delta\varepsilon_{\mathrm{rad2}}$
				& $\propto\delta\varepsilon_{\mathrm{rad2}}$
					& cosmological constant
						& 0
 \\ \hline
	10.
		& $w=-1$
			& $\gamma=\frac{1}{3}$ (DE$ \rightarrow $DM)
				& $\propto\varphi$
				& $\propto\varphi$
				& $\propto\varphi$
					& cosmological constant
						& const$\times\varphi$
 \\
\bottomrule
\end{tabularx}
\caption{\label{results_table_II} Values of the parameters that are compatible with the mechanical approach for the model with non-linear interaction between DM and DE in the case $w+\gamma<0$, as derived in Subsection \ref{sec3.2}.}
\end{table}

On the present paper we have considered a cosmological model with an interaction in the dark sector. The interaction between dark matter (DM) and dark energy (DE) is non-linear and of the form given in \rf{2.11}. The Universe has been investigated at the late stage of its evolution, when the peculiar velocities of the inhomogeneities in the form of galaxies and groups of galaxies are very non-relativistic. We have also investigated the possibility that the rest of the matter sources (e.g. radiation, DM and DE) are in a very specific "coupled" form where their peculiar velocities are also non-relativistic. Such type of fluids were considered, e.g., in \cite{coupled,BUZ1,Laslo2,CPL,Chaplygin,scal field}. This approach is called the mechanical one \cite{EZcosm2,EZcosm1}. From the physical point of view, this happens very naturally when the fluctuations of these matter sources are concentrated around galaxies and groups of galaxies. All the cases consistent within the mechanical approach are summarized in Tables~\ref{results_table_I} and \ref{results_table_II}, which are organized according to the values of the EoS parameter of DE, $w$, and of the interaction parameter, $\gamma$. When $w+\gamma=0$, we have obtained five possible cases that are compatible with the mechanical approach. Of these cases, only two can describe the late-time acceleration of the universe: one corresponding to a cosmological constant without interaction, i.e., $w=-1$ and $\epsdmnow=0$, which is of no interest for us; and the second one corresponding to phantom dark energy behaviour, with $w<-1$. The later solution is of interest because for $w+\gamma=0$ the ratio between the energy densities of DM and DE is constant and therefore there is no coincidence problem in this scenario. However, as we can see from Eq.~\rf{1.12a}, this ratio $\xi_0$ is very small, $\xi_0 << 1$, for the EoS parameter $w^* \sim -1$ dictated by observations (see, e.g., \cite{observ rectr}). As such, in order to get $\overline\varepsilon_{\mathrm{DM}}/\overline\varepsilon_{\mathrm{DE}}\sim O(1)$, we need to include in the model a sufficient fraction of non-interacting DM. Obviously, this does not solve or alleviate the coincidence problem.

When $w+\gamma<0$, we have obtained ten different cases. Of all these possibilities, only four cases (see lines 7-10 in the Table~\ref{results_table_II}) can give rise to an explanation for the the late-time acceleration, of which only one (line 7) does not corresponds asymptotically to the cosmological constant.

In these tables, we also describe the form of the fluctuations ${\delepsrad}_2, \delepsde$ and $\delepsdm$. For each case (i.e. for each pair of the parameters $(w,\gamma)$ in the tables), the form of ${\delepsrad}_2$ is defined in the previous section. With respect to $\delepsde$ and $\delepsdm$, we should use the formulas \rf{1.1} and \rf{1.7}, disregarding any terms that are outside of the accuracy of the mechanical approach. For example, in the line 10 of the Table \ref{results_table_II}, corresponding to the pair $
w=-1$ and $\gamma=1/3$, we get:
\be{4.1}
\delepsdm =3\delepsde=3 \epsdenow\left(\frac{1}{1+\xi_0}\right)^{1/2} \left(\frac{a_0}{a}\right)^3\frac{\varphi}{c^2 a_0}+ O(1/a^5)\,.
\ee

Additionally, for each selected case, we describe the behaviour of r.h.s. of Eq.~\rf{1.9}. It can be easily realised that within the mechanical approach accuracy, r.h.s. of this equation is either equal to zero or proportional to the comoving gravitational potential $\varphi (\bf r)$. The later possibility is of special interest because in this case instead of the Poisson equation\footnote{Strictly, we can speak about the Poisson equation only in the case of a spatially flat topology $\mathcal{K}=0$.}, we get the Helmholtz equation:
\be{4.2}
\Delta\varphi +\left(3\mathcal{K}-A\right)\varphi= \frac{\kappa c^4}{2}(\rho_{\mathrm{c}}-\overline\rho_{\mathrm{c}})\,,
\ee
where the constant $A$ is defined for each case separately. For example, in the case corresponding to the line 10 in the Table \ref{results_table_II}, we obtain
\be{4.3}
A=\frac{2\kappa}{\sqrt{1+\xi_0}} a_0^2\epsdenow\,.
\ee
In Eq.~\rf{4.2}, the comoving rest mass density of the inhomogeneities (e.g. galaxies of mass $m_i$) is
\be{4.4}
\rho_{\mathrm{ c}} = \frac{1}{\sqrt{\gamma}}\sum_i m_i \delta(\mathbf{r}-\mathbf{r}_i)\,.
\ee
The solutions of Eq.~\rf{4.2} can be found for any spatial topology (i.e. for any sign of $\mathcal{K}$) \cite{BUZ1},
and they have a Yukawa-type form. For example, in the case of a flat topology ($\mathcal{K}=0$), we get
\be{4.5}
\varphi=-G_N\sum_i\frac{m_i}{|{\bf r}-{\bf r}_i|}\exp\left(-|{\bf r}-{\bf r}_i|/\lambda\right)+ \lambda^2 \, 4\pi G_N\overline\rho_{\mathrm{c}}\,,
\ee
where $\lambda^2 \equiv \left[A-3\mathcal{K}\right]^{-1}=A^{-1}$.
Such a form for the potential can solve different gravitational paradoxes such as the Neumann-Seeliger paradox \cite{EZcosm2,EZcosm1,Norton} where the sum of an infinite number of Newtonian potentials does not converge. Another very important feature of such potentials is that their averaged values over the whole Universe are equal to zero: $\overline\varphi =0$ \cite{BUZ1}.

Obviously, the Yukawa-type for the gravitational potential has arisen due to the screening of galaxies by the fluctuations of the interacting DE and DM. Usually, this takes place in the case when the fluctuations of the interacting DE and DM are proportional to the gravitational potential: $\delepsdm,\delepsde\sim \varphi$ (see, e.g., Eq.~\rf{4.1}). This leads to another physical reasonable result: these fluctuations averaged over the Universe are also equal to zero.
Given that, as obtained in Eq.~\eqref{4.2}, the gravitational potential $\varphi$ is sourced by $\delta\rho_{\mathrm{c}}=\rho_{\mathrm{c}}-\overline\rho_{\mathrm{c}}$, we find that whenever $\delepsdm,\delepsde\sim \varphi$ the fluctuations of DM and DE appear concentrated around the galaxies and groups of galaxies, i.e., the perfect fluids that describe the interacting DM and DE are really coupled to the inhomogeneities.
This happens, e.g., in the cases corresponding to the lines 7 and 10 of Table~\ref{results_table_II}, with the pairs $(w=-2/3,\gamma=1/3)$ and $(w=-1,\gamma=1/3)$, respectively. These two scenarios are the most interesting and promising cases since for both of them the value $\gamma =1/3$ is rather close to the upper observational limit of $\gamma$ at 68\% confidence level, i.e., $\gamma = 0.259$ as found in \cite{observ rectr}. Despite the fact that they have the same value of $\gamma$, the different values of $w$ in these two scenarios lead to very different evolutions for the Universe, each of them with its own advantage. On the one hand, the ratio of the interacting DM and DE energy densities is defined by formula \rf{0.4}. Obviously, the value $w+\gamma=0$ corresponds to the constant ratio and, consequently, to the absence of the coincidence problem. Therefore, the closer the sum $|w+\gamma|$ to zero, the less severe the coincidence problem is. From this point the case with $w=-2/3$ is preferable.
On the other hand, the value $w=-1$ is much more close to the observable restrictions $-0.9999 <w<-0.9586$ \cite{observ rectr}. Nevertheless, it is worth noting that in both of these scenarios the coincidence problem is less severe than in the standard $\Lambda$CDM model with $w=-1$ and $\gamma=0$.

To conclude, we would like to mention that the gravitational potential $\Phi$ obtained in our paper for the considered model depends on time (see Eqs. \rf{3.1a} and \rf{4.5}). The time dependence of $\Phi$ determines the  integrated Sachs-Wolfe (ISW) effect \cite{ISW0}. The decay of the gravitational potential with the scale factor affects the low multipoles of the radiation power spectrum by generating ISW temperature anisotropies. Therefore, the interaction in the dark sector results in a ISW effect (see also  \cite{ISW1,ISW2,ISW3,ISW4,ISW5,ISW6,ISW7}), and the recent observations can be used to get restrictions on the parameters of the considered model. We will tackle this interesting issue for the nonlinear interaction \rf{2.11} in the near future. This will reveal  which combinations of $\omega$ and $\gamma$ influence the predicted values of the inflationary parameters.

\


\section*{Acknowledgements}
The work of MBL is supported by the Portuguese Agency “Funda\c{c}\~ao para a Ci\^encia e Tecnologia” through an Investigador FCT Research contract, with reference IF/01442/2013/ CP1196/CT0001. She also wishes to acknowledge the partial support from the Basque government Grant No. IT592-13 (Spain) and FONDOS FEDER under grant FIS2014-57956-P (Spanish government). This research work is supported by the Portuguese grant UID/MAT/00212/2013.
 J.M. is thankful to UPV/EHU for a PhD fellowship and UBI for hospitality during the completion of part of this work and acknowledges the support from the Basque government Grant No. IT592-13 (Spain) and Fondos FEDER, under grant FIS2014-57956-P (Spanish Government). A.Zh. acknowledges the hospitality of UBI during his visits in 2014 and 2015 during the completion of part of this work.





\end{document}